\documentclass{nature}

\usepackage{graphicx}

\makeatletter
\let\saved@includegraphics\includegraphics
\AtBeginDocument{\let\includegraphics\saved@includegraphics}
\renewenvironment*{figure}{\@float{figure}}{\end@float}
\makeatother
\usepackage{color}

\def\hl#1{\textcolor{blue}{#1}}

\title{
A new generation of subnanometer-sized materials reveals a general surface polarons property 
}

\author{
Mar\'{\i}a Pilar de Lara-Castells*$^{1}$
and Salvador Miret-Art\'{e}s$^{1}$
}
\begin{document}
\maketitle
\begin{affiliations}
 \item Instituto de F\'isica Fundamental,
CSIC, Serrano 123, 28006 Madrid, Spain.
\end{affiliations}
\noindent 
\noindent {\bf *Email}: Pilar.deLara.Castells@csic.es

\begin{abstract}
The recent advent of cutting-edge experimental techniques allows for a precise synthesis of monodisperse subnanometer metal clusters composed by just a few atoms, and opens new possibilities for subnanometer science. 
The decoration of titanium dioxide surfaces with the Ag$_{5}$ atomic cluster enables the stabilization of surface polarons. A new electron polarization phenomenon accompanying surface polaron formation has thus been revealed.
\end{abstract}

\noindent The very recent development of highly selective experimental techniques making possible the synthesis of subnanometer metal clusters is pushing our understanding of these, more ``molecular" than ``metallic", systems far beyond the present knowledge in materials science.
Subnanometer metal clusters with sizes below 10$-$15 \AA{} (i.e., approximately 100$-$150 atoms) have physical and chemical properties differing significantly from those of nanoparticles and bulk materials, due to quantum confinement effects.\cite{Jena-2018,Zhou-2016} When the cluster size is reduced to a very small (less than 10) number of atoms, the $d$-band of the metal splits into a subnanometer-sized network of discrete molecule-like $d$ orbitals, with the inter-connections having the length of a chemical bond (1$-$2 \AA{}). The spatial structures of the molecular orbitals of these clusters make all the metal atoms cooperatively active and accessible, leading to the appearance of their novel properties and air-stability
(see, e.g., Refs.~\citenum{Conception-2017,arturo-cu5,Alex-2019,PAPER} and references therein). Their possible applications range from visible-light photo-activation of semiconductor surfaces\cite{pilar-agn-tio2,ref18-patricia} to innovative drugs for cancer therapies.\cite{d} For instance, new catalytic and optical properties are acquired by titanium dioxide (TiO$_{2}$) surfaces when decorated with the Cu$_{5}$ atomic cluster:\cite{ref18-patricia,Patricia-2019} The Cu$_{5}$ cluster shifts the absorption of sunlight towards visible light, where the sun emits most of its energy.\cite{ref18-patricia} The coated titanium dioxide stores the absorbed energy temporarily in the form of charge pairs, i.e., electrons and holes, in the direct vicinity of 
the surface, which is a perfect prerequisite for follow-up chemistry.\cite{Yates95}

\noindent Seeking further applications, we have recently modified the TiO$_{2}$ surface with the  atomic silver cluster Ag$_{5}$.\cite{PAPER}
In this way, we have found that TiO$_{2}$-modified surfaces are both visible-light photo-active materials and potential photocatalysts for CO$_{2}$ reduction.\cite{PAPER} Most importantly, we have uncovered new relevant fundamental insights into a general polarization phenomenon accompanying the formation of surface polarons issued from the state-of-the-art of ab-initio calculations and not from a model Hamiltonian,\cite{PAPER} which is summarized in this contribution.

\section{Surface polaron formation and Ag$_{5}$--induced stabilization}

The polaron concept, firstly proposed by Landau (see Ref.~\citenum{Landau-1933}), characterizes an electron moving in a dielectric crystal such as titanium dioxide (TiO$_{2}$). Defects of TiO$_{2}$ surfaces such as oxygen vacancies lead to excess electrons which become self-trapped in Ti$^{3+}$ 3$d^{1}$ states, as 
illustrated in \hl{panel~(a) of Figure 1}. This feature has been predicted theoretically and probed experimentally (for a comprehensive overview see Ref.~\citenum{PRB-polarons-2018}).
In order to screen a self-trapped Ti$^{3+}$ 3$d^{1}$ electron, the O$^{2-}$ anions depart from their equilibrium positions, as indicated with red arrows in \hl{panel (b) of Figure 1}. This lattice distortion is known as the phonon cloud, and the entity formed by the electron and its associated phonon cloud is the polaron. A very simplified picture of the polaron is illustrated in \hl{panel (c) of Figure 1}. Furthermore, the electron also carries a polarization cloud which modifies the electronic structure in its vicinity, characterizing a polarization phenomenon associated with the formation of a surface polaron.\cite{PAPER}

\begin{figure}
\includegraphics[width=\columnwidth]{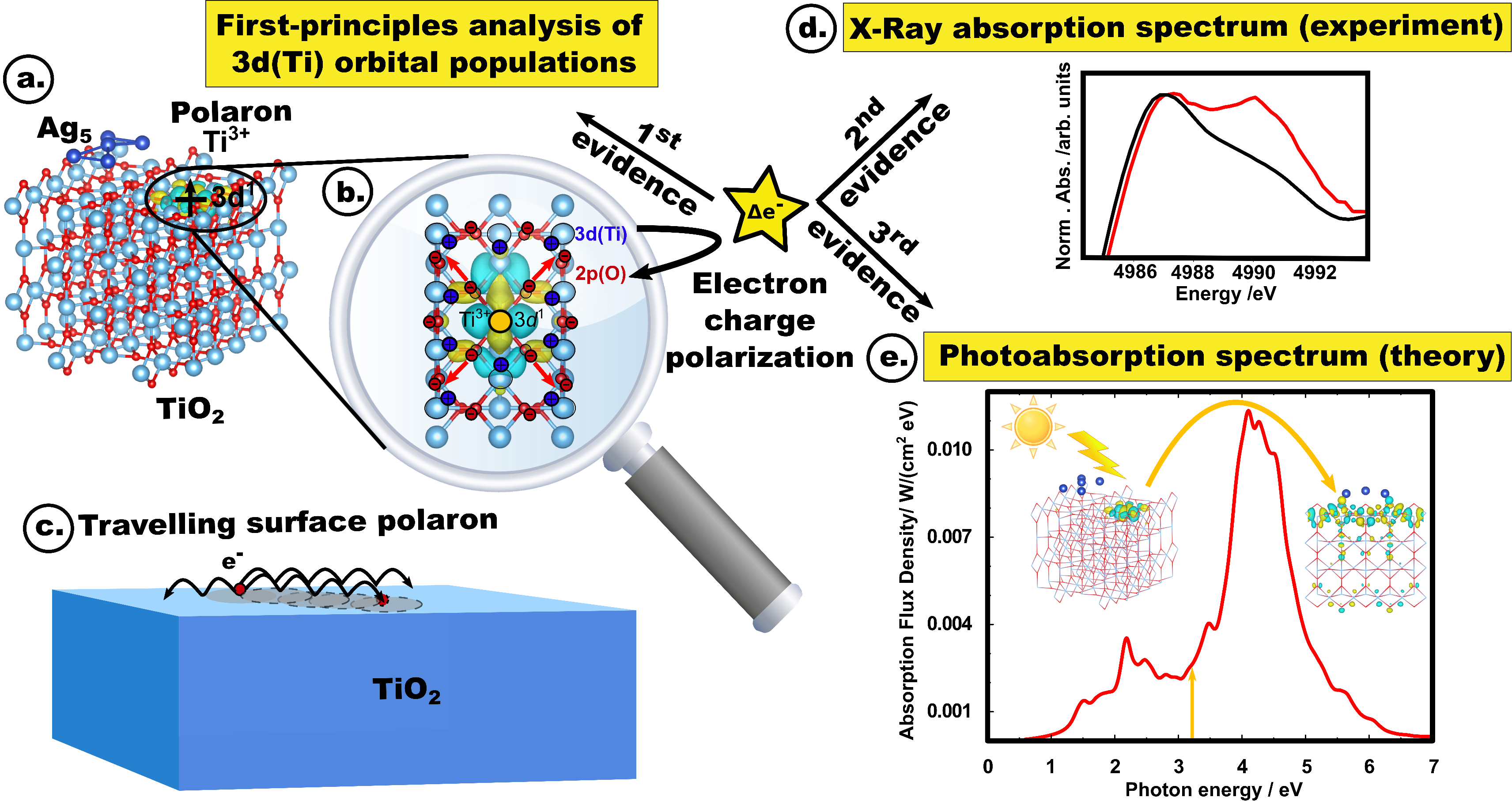}
\caption{{\bf a}) Figure illustrating the formation of a surface polaron as a self-trapped 3$d^{1}$ electron. {\bf b}) Figure illustrating how a self-trapped Ti$^{3+}$ 3$d^{1}$ electron repels nearby oxygen ions while attracting nearby titanium cations, which, in turn, affects the local electronic structure. {\bf c}) Simplified view of a surface polaron. {\bf d)} X-ray absorption spectrum at the Ti K-edge of bare TiO$_{2}$ (black line) and Ag$_{5}$-modified TiO$_{2}$ nanoparticles (red line). {\bf e)} Theoretical UV-Vis spectra showing how the polaronic Ti$^{3+}$ 3$d^{1}$ state is modified when exciting it with a photon energy of 
about 3.1 eV (marked with a yellow arrow).\cite{PAPER}
}
\end{figure}

\noindent Surface Ti$^{3+}$ cations hosting polarons are easily oxidized so that 
their formation is more typical at subsurface Ti$^{3+}$ layers.\cite{PRB-polarons-2018} It was found that a way to stabilize surface polarons on the TiO$_{2}$ surface is via 
the deposition of subnanometer silver (Ag$_{5}$) clusters.\cite{PAPER} The first step of the  
Ag$_{5}$--induced mechanism for the surface polaron formation is the donation of the Ag$_{5}$ unpaired electron to the TiO$_{2}$ surface. As shown in \hl{panel~(a) of Figure~1}, the donated electron becomes localized in one specific 3$d$ orbital lying at the surface, centered at the Ti atom right below the Ag$_{5}$ cluster. After receiving this extra electron, the Ti$^{4+}$ cation becomes formally a Ti$^{3+}$ cation. As indicated with red arrows in \hl{panel~(b) of Figure~1}, the formation of this Ti$^{3+}$ 3$d^{1}$ polaronic state is correlated with the outward movement of the neighboring oxygen anions, i.e., with the typical lattice deformation accompanying the 
formation of polarons. Moreover, an attractive electrostatic interaction between the localized Ti$^{3+}$ 3$d^{1}$ electron and the positively charged Ag$_{5}$ cluster favours the stabilization of the Ag$_{5}$--induced surface polaron.

\section{Evidence of the surface polaron-induced electron polarization phenomenon}

\noindent As illustrated in Figure~1, the evidence for the electron polarization phenomenon 
has been found in three different ways as follows:\cite{PAPER} 

\noindent (1) Our first-principles analysis has provided evidence for the Ag$_{5}$--induced depopulation of 3$d$(Ti) orbitals, typically in favour of 2$p$(O) orbitals, despite of the fact that polarons are characterized by an excess charge (i.e., ``extra" electrons) trapped at the Ti sites hosting them. In fact, the opposite behaviour is found through the first-principles modelling of the non-polaronic counterpart, already indicating that the depopulation of 3$d$(Ti) orbitals is a feature characterizing the polaronic state.

\noindent (2) A second evidence of the depopulation of 3$d$(Ti) orbitals in favour of 2$p$(O) orbitals has been found through X-ray absorption near edge structure (XANES) spectroscopy. This technique is characterized by a high chemical selectivity. \hl{Panel~(d) of Figure 1} shows the XANES spectra at the Ti $K$-edge of bare TiO$_{2}$ nanoparticles (shown in black) and Ag$_{5}$--modified TiO$_{2}$ nanoparticles (shown in red). The increase of the spectral feature appearing at ca. 4990 eV for Ag$_{2}$--modified TiO$_{2}$ nanoparticles indicates a charge transfer from Ti cations to O anions. It gives rise to an average decrease of the 3$d$(Ti) orbitals population.

\noindent (3) Moreover, a first-principles simulation of the UV-Vis optical response of the surface polaron provided further insights into the correlation between the depopulation of the 3$d$(Ti) orbitals and the polaron formation as follows: the excitation of the Ag$_{5}$--TiO$_{2}$ system with a photon energy at the end of the visible region causes the ``jump" of the polaron Ti$^{3+}$ 3$d^{1}$ electron to the 3$d$ orbitals of the surface Ti atoms which have suffered depopulation. Hence, the polaron induces a hole (lack of one electron) which is extended over the surface Ti atoms in its vicinity and this hole becomes filled upon photo-excitation 
of the polaron. The photo-absorption spectrum is illustrated in \hl{panel~(e) of Figure 1}, together with the jumping of an electron from the polaron Ti$^{3+}$ 3$d^{1}$ state to an acceptor state formed by many surface $3d$(Ti) orbitals. 

\noindent The responsible mechanism for the average depopulation of 3$d$(Ti) orbitals in favour of 2$p$(O) orbitals in the presence of a surface polaron can be explained as follows: the self-trapped Ti$^{3+}$ 3$d^{1}$ electron repels nearby oxygen anions and attracts nearby titanium cations which, in turn, affects their electronic structure, causing the transfer of electronic charge from Ti$^{4+}$ cations to O$^{2-}$ anions (see \hl{panel (b) of Figure 1}). This mechanism can be considered as a new electron polarization phenomenon of surface polarons which is not just associated with Ag$_{5}$-induced surface polarons: the same conclusions are reached when analysing the polarons formed from oxygen vacancies in a reduced TiO$_{2}$ surface.\cite{PAPER} 

Thus, the TiO$_{2}$ surface modification with the subnanometer-sized silver cluster Ag$_{5}$ has served to uncover a new way of stabilizing surface polarons.\cite{PAPER} Furthermore, a novel polarization phenomenon has been revealed, which arises from quantum effects introduced by the adsorbed atomic Ag$_{5}$ cluster. This phenomenon 
has its origin in the subnanometer size of the actual quantum system composed by just 5 atoms and is therefore fundamentally different from phenomena known for larger clusters on the nanoscale or the bulk. The exploration of the internal structure of the polarons has been achieved by combining state-of-the-art first principles theory with X-ray absorption near edge structure (XANES) spectroscopic measurements. Remarkably, the surface polaron property has been deduced theoretically not from a model Hamiltonian but from extended state-of-the-art {\it ab initio} calculations.
This combined theoretical-experimental approach might be useful for future studies of surface polaron interactions, a matter 
of high relevance at present.\cite{nature-materials-polaron-2018} It is also expected that our findings will not only contribute to an improved fundamental understanding of surface polarons but also to their controlled formation and usage in applications. Our very recent recent studies\cite{PAPER,pilar-agn-tio2,ref18-patricia,Patricia-2019,Alex-2019} thus confirm the great potential that lies in these new class of materials, which are shaping the modern field of subnanometer science.

\section{Bibliography}

\begin{addendum}

\item[Acknowledgement] 
This work has been partly supported by the Spanish Agencia Estatal de Investigaci\'{o}n (AEI) and the Fondo Europeo de Desarrollo Regional (FEDER, UE) under Grant No. MAT2016-75354-P. 
\item[Mar\'{i}a Pilar de Lara-Castells] 
currently has a permanent researcher position 
(group leader) at the Institute of Fundamental Physics (IFF) in the Spanish National Research Council (CSIC), leading the AbinitSim Unit and being 
principal investigator of the project NANOABINIT 
(Ref.~MAT2016-75354-P). She conceptualized and coordinated the joint theoretical-experimental work outlined in this feature.
\item[Salvador Miret-Art\'{e}s] is currently full professor and director of the Institute of Fundamental Physics (IFF) in the Spanish National Research Council (CSIC).
\end{addendum}
\includegraphics[width=0.22\textwidth]{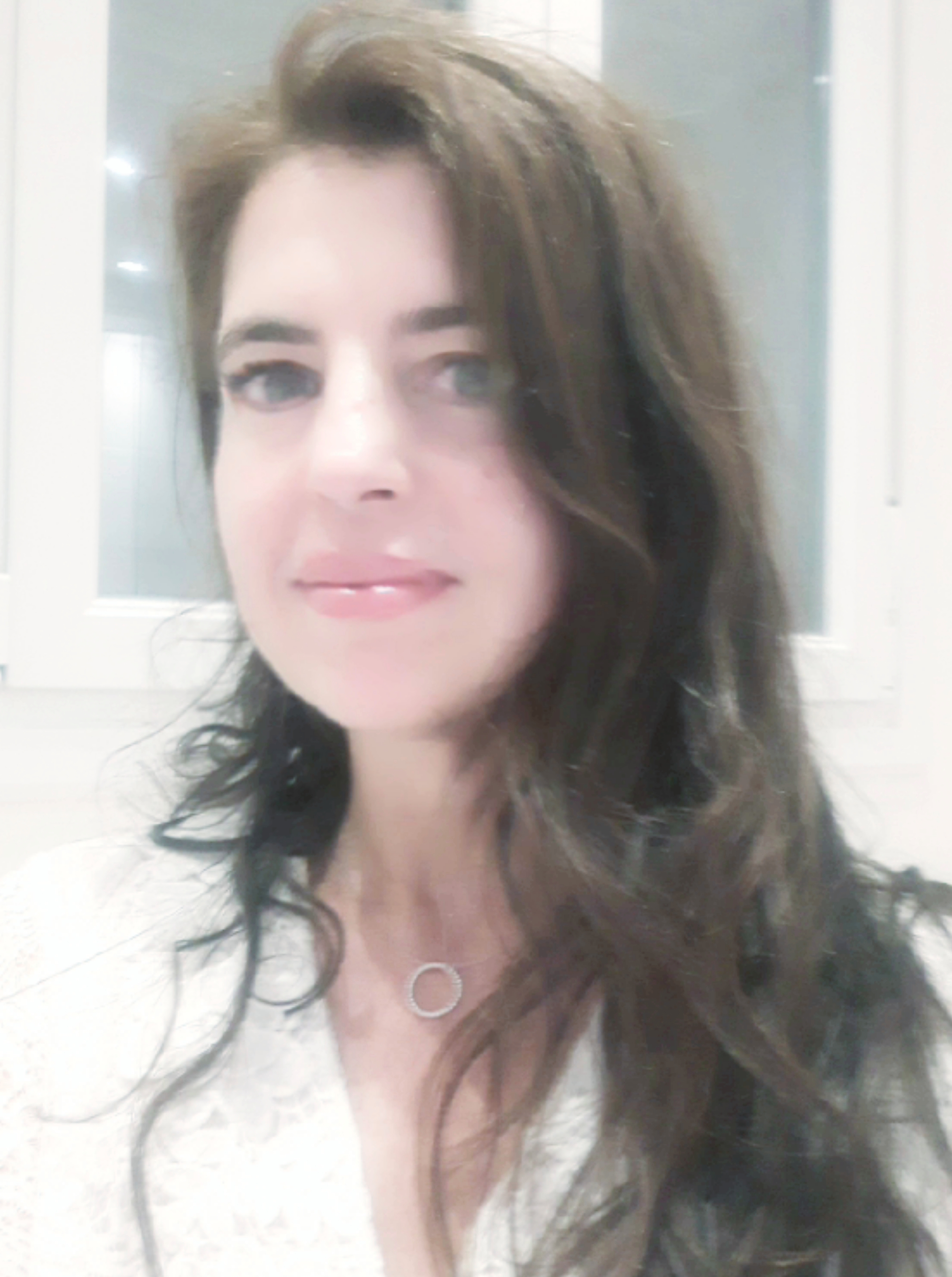}
\includegraphics[width=0.22\textwidth]{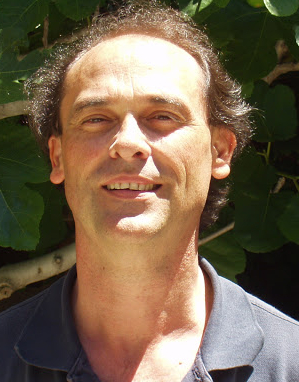}

\end{document}